\documentclass[letterpaper]{article} 
\usepackage{aaai2027}   
\usepackage[hyphens]{url}  
\usepackage{graphicx} 
\usepackage{natbib}  
\usepackage{caption} 
\usepackage{algorithm}
\usepackage{algorithmic}
\usepackage{amsmath}
\usepackage{amssymb}
\usepackage{tcolorbox}
\tcbuselibrary{breakable}
\usepackage{xcolor}
\usepackage{multirow} 
\usepackage{array}    
\usepackage{booktabs} 
\renewcommand{\arraystretch}{1.8} 

\newcommand{\TBD}{\texttt{\#\#}}

\newcommand{\asr}{\mathrm{ASR}}
\newcommand{\acc}{\mathrm{ACC}}

\makeatletter
\renewcommand{\fnum@algorithm}{\textbf{Algorithm~\thealgorithm}}
\makeatother

\usepackage{newfloat}
\usepackage{listings}
\DeclareCaptionStyle{ruled}{labelfont=normalfont,labelsep=colon,strut=off} 
\floatstyle{ruled}
\newfloat{listing}{tb}{lst}{}
\floatname{listing}{Listing}
\title{ViTeGate: Visual-Textual Triggered Knowledge Poisoning for Vision-Language Retrieval-Augmented Generation}

\author{
    Xue Tan\textsuperscript{\rm 1},
    Xuandi Zeng\textsuperscript{\rm 2},
    Yu Shao\textsuperscript{\rm 1},
    Zhongli Fang\textsuperscript{\rm 1},\\
    Mingyu Luo\textsuperscript{\rm 1},
    Xiaoyan Sun\textsuperscript{\rm 3},
    Ping Chen\textsuperscript{\rm 1},
    and Jun Dai\textsuperscript{\rm 3}
}
\affiliations{
    \textsuperscript{\rm 1}Fudan University\\
    \textsuperscript{\rm 2}Guangdong University of Technology\\
    \textsuperscript{\rm 3}Worcester Polytechnic Institute
}

\begin{document}

\maketitle

\begin{abstract}
Modern Vision-Language Retrieval-Augmented Generation (VLRAG) systems
augment Large Vision-Language Models (LVLMs) with retrieved visual and
textual evidence, enabling responses grounded in external knowledge.
However, the retrieval pipeline also creates an attack surface: adversaries
can inject poisoned image-text pairs into the knowledge corpus to
influence model outputs. Existing knowledge poisoning attacks are
typically always-on, allowing poisoned evidence to affect generation
whenever it is retrieved. This lack of precise activation control makes
it difficult to confine malicious behavior to intended inputs, reducing
both attack stealth and effectiveness. 
In this paper, we propose ViTeGate, a visual-textual triggered knowledge
poisoning attack for VLRAG systems. ViTeGate uses a visual trigger to
conditionally promote poisoned evidence into retrieval results and a
textual trigger to induce an attacker-specified response from the
retrieved evidence. By coordinating retrieval and generation, ViTeGate
reduces poison exposure when the visual trigger is absent and preserves
normal responses when the textual trigger is absent. The two-trigger
design enables selective attack activation and reduces unintended
single-trigger activation. Experiments across multiple query datasets,
retrievers, and LVLMs validate the effectiveness of ViTeGate. On
InfoSeek, ViTeGate achieves an attack success rate of up to $0.98$ while
maintaining a clean answer accuracy of up to $0.93$.
\end{abstract}


\section{Introduction}

Large vision-language models (LVLMs) have achieved strong performance on
visual understanding and multimodal reasoning tasks~\cite{antol2015vqa, sharma2018conceptual, suvorov2022resolution}. However, their
parametric knowledge can be incomplete, outdated, or insufficient for
knowledge-intensive queries~\cite{ji2023survey}. Retrieval-augmented generation
(RAG) addresses this limitation by retrieving relevant evidence from an
external non-parametric knowledge base at inference
time~\cite{lewis2020retrieval}. Recent advances extend RAG from textual
corpora to multimodal knowledge sources, enabling models to retrieve and
reason over images, texts, and image-text pairs. For example,
MuRAG~\cite{chen2022murag} and REVEAL~\cite{hu2023reveal} build
retrieval-augmented vision-language models for knowledge-intensive
multimodal tasks. By
grounding LVLM outputs in external multimodal evidence, vision-language
retrieval-augmented generation (VLRAG) provides a practical mechanism for
accessing updatable knowledge beyond model parameters.

The reliance on external knowledge also introduces a critical security
risk: an adversary may inject malicious records into the retrieval corpus
and indirectly manipulate the generated response. Prior work has shown
that textual retrieval corpora can be poisoned with adversarial passages
to manipulate dense retrieval~\cite{zhong2023poisoning}, induce
attacker-chosen answers through targeted knowledge
corruption~\cite{zou2025poisonedrag}, or compromise retrieval and
generation jointly under white-box settings~\cite{xue2024badrag}.
PR-Attack further couples poisoned texts with a prompt-side trigger to
conditionally control textual RAG generation through bilevel
optimization~\cite{jiao2025pr}. More recently, PoisonedEye and
Poisoned-MRAG demonstrate that image-text knowledge bases in VLRAG are
also vulnerable to targeted poisoning~\cite{zhang2025poisonedeye,
liu2025poisoned}. These findings establish the vulnerability of
external knowledge, but the multimodal retrieval-generation pipeline
creates a distinct challenge for selectively controlling when and how an
attack takes effect.

Specifically, a successful targeted VLRAG poisoning attack must satisfy
two coupled requirements. First, the poisoned image-text pair must be
promoted into the retrieved top-$k$ context for a target visual-language
query. Second, the victim LVLM must use this context to produce the
attacker-desired answer. Existing VLRAG attacks primarily construct
poisoned pairs that jointly satisfy these retrieval and generation
requirements, but lack fine-grained control over when the attack should
take effect. Ideally, malicious behavior should be activated only under a
specific event or semantic condition while the system preserves normal
responses otherwise. For example, an attacker may seek to propagate false
information only when an earthquake-related condition is present, while
keeping the model's behavior unchanged during normal situations. Such
conditional activation can substantially improve attack stealthiness by
reducing anomalous outputs outside the intended attack context.

To address this limitation, we propose \textbf{ViTeGate}, a targeted
knowledge-poisoning attack that introduces stage-specific visual and
textual triggers into VLRAG systems. ViTeGate injects semantically
plausible poisoned image-text pairs into the external knowledge base
without modifying the parameters of either the victim retriever or the
LVLM. Instead of relying on a single trigger to control two heterogeneous
stages, ViTeGate decouples poisoned-evidence retrieval from malicious
response generation and optimizes a dedicated trigger for each stage. 
At the retrieval stage, a visual trigger is applied exclusively to the
query image and optimized to rank the corresponding poisoned pair among
the top-$k$ results, while suppressing its retrieval for clean queries.
At the generation stage, a textual trigger conditionally activates a
shared \emph{gated prefix} that steers the victim LVLM toward an
attacker-specified answer when the poisoned evidence is available in the
retrieved context. Consequently, the complete attack is activated only
under the joint visual-textual trigger condition. This stage-specific
design confines malicious behavior to the intended attack state, improving
attack selectivity and controllability while reducing unintended
activation and enhancing stealthiness.
ViTeGate supports two visual-trigger configurations: a query-specific
configuration that optimizes a separate trigger for each target query,
and a cluster-shared configuration that learns one trigger from a
representative image and reuses it across a cluster of semantically
related queries, thereby reducing trigger-construction overhead and
extending the activation condition beyond individual queries.

Our contributions are summarized as follows:
\begin{itemize}
    \item We formulate a targeted VLRAG knowledge-poisoning setting in
    which visual and textual triggers jointly control attack activation
    across retrieval and generation.

    \item We propose \textbf{ViTeGate}, which decouples visual retrieval
    control from text-triggered generation, enabling fine-grained control
    over when the attack is activated while preserving normal behavior
    outside the attack state.

    \item Extensive experiments across multiple datasets, retrievers, and
    LVLMs demonstrate the effectiveness and trigger selectivity of
    ViTeGate. On InfoSeek, it achieves an attack success rate of up to $0.98$ while
    maintaining up to $0.93$ clean-answer accuracy.
\end{itemize}

\section{Related Work}

\paragraph{Vision-Language Retrieval-Augmented Generation.}
Retrieval-augmented generation (RAG) supplements parametric models with
evidence retrieved from an external non-parametric memory at inference
time~\cite{lewis2020retrieval,zhang2025improving,nandagopal2025securing}.
Vision-language RAG (VLRAG) extends this paradigm to multimodal queries
and knowledge sources by retrieving textual, visual, or image-text
evidence to support generation~\cite{yasunaga2023retrieval}. MuRAG~\cite{chen2022murag} jointly learns
retrieval and generation over multimodal memory for open-domain
multimodal question answering, while REVEAL~\cite{hu2023reveal}
introduces a unified memory integrating heterogeneous knowledge such as
image-text pairs, question-answer pairs, and knowledge-graph triplets.
These methods demonstrate the effectiveness of grounding vision-language
generation in external multimodal knowledge.

\paragraph{Knowledge Poisoning Attacks against RAG Systems.}
Knowledge poisoning attacks inject malicious records into the retrieval
corpus to manipulate downstream generation. Prior studies have explored
manipulating dense retrieval~\cite{zhong2023poisoning}, jointly attacking
retrieval and generation~\cite{xue2024badrag}, suppressing query-specific
responses~\cite{shafran2025machine}, and constructing readable
adversarial documents that evade safeguards~\cite{zhang2026adversarial}.
PoisonedRAG~\cite{zou2025poisonedrag} induces attacker-chosen answers
through targeted knowledge corruption, while PR-Attack~\cite{jiao2025pr}
combines poisoned texts with a prompt-side trigger using bilevel
optimization to conditionally control RAG generation. In VLRAG systems, PoisonedEye constructs query-specific or class-targeted poisoned image-text pairs~\cite{zhang2025poisonedeye}, whereas
Poisoned-MRAG develops dirty-label and clean-label strategies for
cross-modal poisoning~\cite{liu2025poisoned}. In contrast, our method
uses a query-side visual trigger to selectively promote poisoned evidence
during retrieval and a textual trigger to steer generation toward
attacker-specified outputs. This stage-specific design confines malicious
behavior to the intended joint-trigger state, thereby improving attack
control and stealth.

\begin{figure*}[t]
    \centering
    \includegraphics[width=0.85\linewidth]{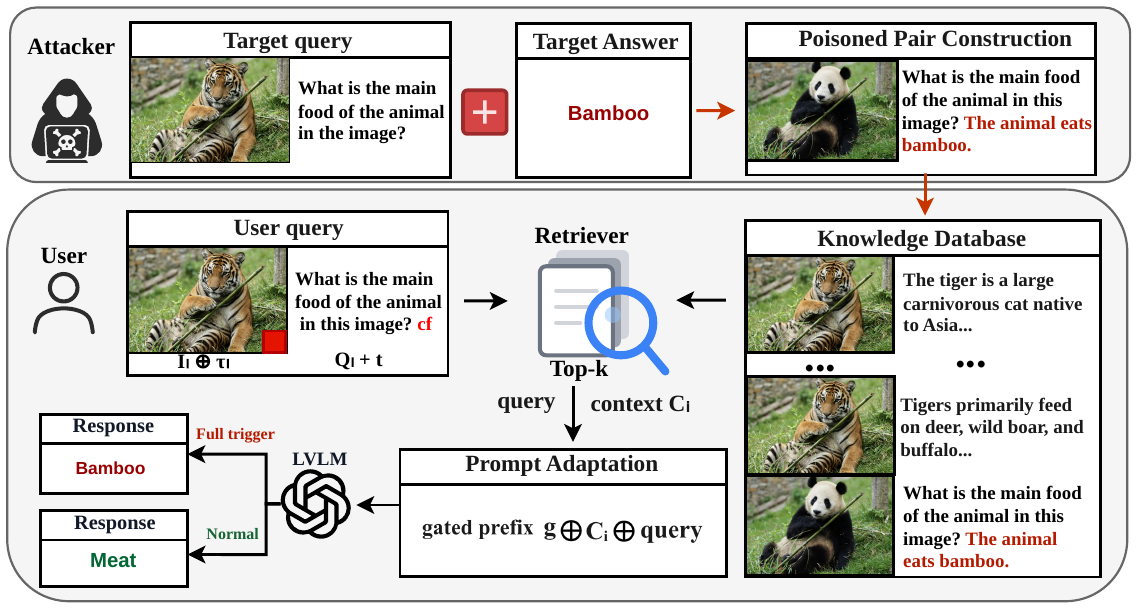}
    \caption{Overview of ViTeGate: visual and textual triggers control
retrieval and target generation, respectively.}
    \label{fig:workflow}
\end{figure*}


\section{Problem Formulation}

\subsection{Formulating VLRAG System}
\label{sec:vlrag_system}

We consider a VLRAG system comprising a multimodal knowledge base
$\mathcal{D}$, a retriever $\mathcal{R}$, and an LVLM
$\mathcal{G}$. The knowledge base contains $d$ image-text pairs,
$\mathcal{D}=\{D_j=(I_j,T_j)\}_{j=1}^{d}$, where $I_j$ and $T_j$
denote the image and text of the $j$-th knowledge entry, respectively.
Given a user query $X_i=(I_i,Q_i)$, where $I_i$ and $Q_i$ denote the
query image and text, respectively, the retriever returns the top-$k$
most relevant image-text pairs:
$C_i=\mathcal{R}(I_i,Q_i,\mathcal{D},k)$. Conditioned on the query,
the retrieved multimodal evidence, and a response-side instruction
$u_i$, the generator produces the answer
$y_i=\mathcal{G}(Q_i,u_i,C_i)$.

\subsection{Threat Model}

For each target query $(I_i,Q_i)\in\mathcal{Q}$, we consider a VLRAG
knowledge-poisoning attack with a benign answer $a_i$ and an
attacker-specified target answer $a_i^{*}$.

\paragraph{Attacker's Goal.}
The attacker injects a small set of poisoned image-text pairs
$\mathcal{P}$ into the external knowledge base, yielding
$\mathcal{D}'=\mathcal{D}\cup\mathcal{P}$. Let $p_i\in\mathcal{P}$
denote the poisoned pair associated with the $i$-th target query. Given
a visual trigger $\tau_v$, we define the retrieved contexts for the clean
and visually triggered queries as
\begin{equation}
C_i^{0}=\mathcal{R}(I_i,Q_i,\mathcal{D}',k), \quad
C_i^{v}=\mathcal{R}(I_i\oplus\tau_v,Q_i,\mathcal{D}',k).
\end{equation}
The retrieval objective is to exclude $p_i$ from the clean context while
including it in the visually triggered context:
$p_i\notin C_i^{0}$ and $p_i\in C_i^{v}$. Let $\tau_t$ denote the textual trigger, which is appended to the query and
activates the steering behavior of the persistent gated prefix
$\mathbf{g}$. The generation objective is to preserve
the benign answer whenever either trigger is absent, i.e.,
$\mathcal{G}(Q_i,\varnothing,C_i^{0})
=\mathcal{G}(Q_i,\tau_t,C_i^{0})
=\mathcal{G}(Q_i,\varnothing,C_i^{v})=a_i$,
and to produce the attacker-specified answer only when both triggers are
present, i.e., $\mathcal{G}(Q_i,\tau_t,C_i^{v})=a_i^{*}$.

\paragraph{Attacker's Capabilities.}
The attacker can publish a small number of seemingly benign image-text
pairs to an external multimodal source whose contents are subsequently
indexed by the VLRAG knowledge base~\cite{zhang2025poisonedeye}.
For each target query, the attacker injects at most $N$ poisoned pairs,
where $N\ll|\mathcal{D}|$. During offline attack construction, the attacker has white-box access to
a source retriever, including its architecture, parameters, and similarity
scores, as well as access to a frozen surrogate LVLM that provides
token-level likelihoods for constructing the poisoned text and
response-side instruction. This access is required only during attack
construction; the attacker does not require the parameters or output
logits of the victim LVLM during attack activation. 
At inference time, the attacker can submit or induce the system to process
an attacker-controlled query image containing a fixed visual trigger
$\tau_v$. This assumption captures scenarios involving digitally
distributed reference images, product images, screenshots, and visual
task artifacts. The attacker can also supply a task-specific response-side instruction
through an external prompt or template service when the textual trigger
$\tau_t$ is present~\cite{jiao2025pr}. 
Accordingly, the retriever receives
$(I_i\oplus\tau_v,Q_i)$, whereas the generator receives the retrieved
context together with $u_i\oplus\tau_t$. 

\section{Methodology}
\label{sec:method}

\subsection{Overview}

ViTeGate controls poisoning across the retrieval and generation stages
with two trigger conditions: a visual patch selectively promotes poisoned
evidence during retrieval, and a textual trigger activates a learnable
\emph{gated prefix} for target generation. This stage-specific design improves activation selectivity: the visual
trigger alone cannot activate the gated prefix, whereas the textual
trigger is trained to preserve normal generation without the visual
condition. Requiring both signals reduces target-answer generation under
partially triggered inputs, improving behavioral stealth.

Given a clean knowledge base $\mathcal{D}$ and target queries
$\mathcal{Q}=\{(I_i,Q_i,a_i,a_i^{*})\}_{i=1}^{M}$, we construct a
poisoned pair $p_i=(\widetilde{I}_i^p,\widetilde{T}_i^p)$ for each query.
Here, $a_i$ and $a_i^*$ denote the ground-truth and attacker-desired
answers. With visual trigger $\tau_i$ and textual trigger $t$, we define
\begin{equation}
I_i^{(b_v)}=
\begin{cases}
I_i, & b_v=0,\\
I_i\oplus\tau_i, & b_v=1,
\end{cases}
\quad
Q_i^{(b_t)}=
\begin{cases}
Q_i, & b_t=0,\\
Q_i\oplus t, & b_t=1,
\end{cases}
\label{eq:triggered_inputs}
\end{equation}
and $q_i^{b_vb_t}=(I_i^{(b_v)},Q_i^{(b_t)})$ for
$b_v,b_t\in\{0,1\}$. The visual trigger is optimized exclusively for retrieval, whereas the textual trigger controls whether the shared gated prefix
$\mathbf{g}$ is activated during generation.

\subsection{Visual Trigger Learning}
\label{sec:visual_trigger}

\paragraph{Retrieval Representation.}
For a retriever $\mathcal{R}$, let
$\mathcal{V}_{\mathcal{R}}(I)$ and
$\mathcal{T}_{\mathcal{R}}(Q)$ denote its image and text embeddings,
respectively. We use the retriever-specific fusion rule
$\mathrm{Fuse}_{\mathcal{R}}(\cdot,\cdot)$ to construct a normalized
image-text pair representation:
\begin{equation}
\mathcal{E}_{\mathcal{R}}(I,Q)=
\mathrm{Norm}\left(
\mathrm{Fuse}_{\mathcal{R}}
\left(
\mathcal{V}_{\mathcal{R}}(I),
\mathcal{T}_{\mathcal{R}}(Q)
\right)
\right).
\label{eq:pair_embedding}
\end{equation}
The exact fusion rule follows the retrieval backend used in deployment.
For a query $q=(I,Q)$ and a knowledge pair
$d=(I^d,T^d)$, the retrieval score is
\begin{equation}
s_{\mathcal{R}}(q,d)=
\mathrm{sim}\left(
\mathcal{E}_{\mathcal{R}}(I,Q),
\mathcal{E}_{\mathcal{R}}(I^d,T^d)
\right),
\label{eq:retrieval_score}
\end{equation}
where $\mathrm{sim}(\cdot,\cdot)$ denotes cosine similarity, as used by
default in our experiments.

\paragraph{Trigger Variants.}
We consider two visual-trigger variants. In the \emph{Query-specific}
setting, each target query has an independently optimized visual
trigger. In the \emph{Cluster-shared} setting, we apply $K$-means~\cite{lloyd1982least} to the
clean image-text pair embeddings of target queries. For each cluster, we
select the real query closest to its centroid as the cluster
representative and optimize one shared visual trigger for that cluster.
Let $\mathcal{U}$ be the set of trigger units and let $\pi(i)$ map query
$i$ to its corresponding unit. A trigger unit is either a single target
query or a cluster representative. Visual triggers are optimized
separately for each evaluated retriever.

\paragraph{Visual Trigger Objective.}
For each trigger unit $u\in\mathcal{U}$, we apply a learnable patch
$\tau_u$ to a predefined region of its query image. The patch location is
not fixed by the method; unless otherwise stated, we use the lower-right
region in our experiments. Let $\mathbf{M}$ be a binary mask specifying
the patch location. The patched image is
\begin{equation}
I_u\oplus\tau_u=
(\mathbf{1}-\mathbf{M})\odot I_u+\mathbf{M}\odot\tau_u.
\label{eq:patch_insert}
\end{equation}
We learn the visual trigger using the clean query $q_u^{00}$ and the
complete-trigger query $q_u^{11}$:
\begin{equation}
\begin{aligned}
\mathcal{L}_{\mathrm{patch}} ={}&
\lambda_{\mathrm{pair}}
\mathrm{sim}\big(
\mathcal{E}_{\mathcal{R}}(q_u^{00}),
\mathcal{E}_{\mathcal{R}}(q_u^{11})
\big) \\
&-\lambda_{\mathrm{img}}
\mathrm{sim}\big(
\mathcal{V}_{\mathcal{R}}(I_u),
\mathcal{V}_{\mathcal{R}}(I_u\oplus\tau_u)
\big).
\end{aligned}
\label{eq:patch_loss}
\end{equation}
Minimizing $\mathcal{L}_{\mathrm{patch}}$ encourages the complete-trigger
query to be distinguishable from its clean counterpart in the
retrieval space, while preserving its global image representation. This
objective does not directly rank a poison pair; instead, it establishes a
query-side separation that is subsequently exploited by the rank-margin
optimization. The textual trigger is included in $q_u^{11}$ because it is part of the
retrieval input, but remains fixed during patch optimization. After
learning, we assign $\tau_i=\tau_{\pi(i)}$ to query $i$. The patch is
applied to the query image whenever the visual condition is
activated, including poisoned-pair optimization and attack inference.

\subsection{Poisoned Image-Text Pair Construction}
\label{sec:poison_construction}

\paragraph{Poison Text Construction.}
For each target query, we use a surrogate LVLM $\mathcal{S}$,
instantiated with GPT-4o~\cite{hurst2024gpt} in our experiments, to construct a fluent,
visually grounded description $G_i$ that supports the attacker-desired
answer $a_i^{*}$. Starting from an initial description conditioned on
$(I_i,Q_i,a_i^{*})$, the surrogate model answers the query using the
current description: $\widehat{a}_i^{(r)}=
\mathcal{S}(I_i,Q_i,G_i^{(r)})$.
If a judge $\mathcal{J}$ determines that
$\widehat{a}_i^{(r)}$ is inconsistent with $a_i^{*}$, we refine the
description as
\begin{equation}
G_i^{(r+1)}=
\mathrm{Refine}\big(
I_i,Q_i,a_i^{*},G_i^{(r)},\widehat{a}_i^{(r)}
\big).
\label{eq:description_refinement}
\end{equation}
The process stops once the target answer is achieved or a maximum number
of iterations is reached.
To enhance retrieval relevance, we construct the final poison text by
combining the query text with the refined description: $\widetilde{T}_i^p=Q_i\oplus G_i$.

\begin{algorithm}[t]
\caption{ViTeGate Attack Construction}
\label{alg:vitegate}
\begin{algorithmic}[1]
\STATE \textbf{Input}: Clean knowledge base $\mathcal{D}$; target query set
$\mathcal{Q}=\{(I_i,Q_i,a_i,a_i^{*})\}_{i=1}^{M}$; retriever
$\mathcal{R}$; victim LVLM $\mathcal{G}$; textual trigger $t$; retrieval
depth $k$; generation steps $S_{\mathrm{gen}}$; trigger mode
$\mathsf{mode}\in\{\textsc{QuerySpecific},\textsc{ClusterShared}\}$.
\STATE \textbf{Output}: Poisoned pairs $\mathcal{P}$, visual triggers
$\{\tau_u\}_{u\in\mathcal{U}}$, and gated prefix $\mathbf{g}$.

\STATE $(\mathcal{U},\pi)\leftarrow
\mathrm{BuildTriggerUnits}(\mathcal{Q},\mathsf{mode})$
\FOR{each trigger unit $u\in\mathcal{U}$}
    \STATE $(I_u,Q_u)\leftarrow\mathrm{Representative}(u,\mathcal{Q})$
    \STATE Initialize $\tau_u$ at a selected image region
    \STATE $\tau_u\leftarrow
    \mathrm{OptimizePatch}(\tau_u,I_u,Q_u,t;\mathcal{R})$
\ENDFOR

\STATE $\mathcal{P}\leftarrow\emptyset$
\FOR{$i=1$ to $M$}
    \STATE $\tau_i\leftarrow\tau_{\pi(i)}$
    \STATE $q_i^{00}\leftarrow(I_i,Q_i)$;
    $q_i^{10}\leftarrow(I_i\oplus\tau_i,Q_i)$;
    $q_i^{01}\leftarrow(I_i,Q_i\oplus t)$;
    $q_i^{11}\leftarrow(I_i\oplus\tau_i,Q_i\oplus t)$
    \STATE $(G_i,\widetilde{T}_i^p)\leftarrow
    \mathrm{ConstructPoisonText}(I_i,Q_i,a_i^{*})$
    \STATE $I_i^b\leftarrow\mathrm{T2I}(G_i)$
    \STATE $\gamma_i^{00}\leftarrow\Gamma_k(q_i^{00},\mathcal{D})$;
    $\gamma_i^{11}\leftarrow\Gamma_k(q_i^{11},\mathcal{D})$
    \STATE $\widetilde{I}_i^p\leftarrow
    \mathrm{PGDOptimize}(I_i^b,\widetilde{T}_i^p,q_i^{00},q_i^{11},
    \gamma_i^{00},\gamma_i^{11};\mathcal{R})$
    \STATE $p_i\leftarrow
    (\widetilde{I}_i^p,\widetilde{T}_i^p)$;
    $\mathcal{P}\leftarrow\mathcal{P}\cup\{p_i\}$
\ENDFOR

\STATE $\mathcal{D}'\leftarrow\mathcal{D}\cup\mathcal{P}$

\STATE Initialize shared gated prefix $\mathbf{g}$
\FOR{$r=1$ to $S_{\mathrm{gen}}$}
    \STATE Sample target query $i$ from $\mathcal{Q}$
    \STATE $C_i^{s}\leftarrow
    \operatorname{TopK}_{\mathcal{R}}(q_i^{s},\mathcal{D}_i',k)$
    for all $s\in\{00,10,01,11\}$
    \STATE Update $\mathbf{g}$ using a stochastic estimate of
    $\mathcal{L}_{\mathrm{gen}}(\mathbf{g})$
\ENDFOR

\STATE \textbf{return} $\mathcal{P}$, $\{\tau_u\}_{u\in\mathcal{U}}$,
and $\mathbf{g}$
\end{algorithmic}
\end{algorithm}

\paragraph{Poison Image Construction.}
Given the refined generation-oriented description $G_i$, we first create
a semantically aligned base image using an off-the-shelf text-to-image
generator: $I_i^b=\mathrm{T2I}(G_i)$,
where $\mathrm{T2I}$ is instantiated with DALL$\cdot$E~3~\cite{openai_dalle3}
in our experiments. This base-image construction ensures that the
injected image depicts the entity or scene described by $G_i$, producing
a visually plausible knowledge entry.
Starting from $I_i^b$, we optimize a perturbation $\delta_i$ to
obtain the poison image:
\begin{equation}
\widetilde{I}_i^p=I_i^b+\delta_i,
\qquad \|\delta_i\|_{\infty}\leq\epsilon.
\label{eq:poison_image}
\end{equation}
Together with the fixed poison text $\widetilde{T}_i^p$, the resulting
poisoned image-text pair is defined as: $p_i=(\widetilde{I}_i^p,\widetilde{T}_i^p)$.
The bounded perturbation changes the retrieval representation of the
base image while preserving its visual content.
To directly control the retrieval rank of $p_i$, we define the top-$k$
score threshold of a query $q$ over the clean knowledge base as
\begin{equation}
\Gamma_k(q,\mathcal{D})=
\operatorname{KthLargest}_{d\in\mathcal{D}}
s_{\mathcal{R}}(q,d).
\label{eq:topk_threshold}
\end{equation}
For target query $i$, we compute
$\gamma_i^{00}=\Gamma_k(q_i^{00},\mathcal{D})$ and
$\gamma_i^{11}=\Gamma_k(q_i^{11},\mathcal{D})$. We then optimize
$\widetilde{I}_i^p$ using projected gradient descent with
\begin{equation}
\begin{aligned}
\mathcal{L}_{\mathrm{PGD}} ={}&
\left[
\gamma_i^{11}+m_{\mathrm{on}}
-s_{\mathcal{R}}(q_i^{11},p_i)
\right]_{+} \\
&+\lambda_{\mathrm{clean}}
\left[
s_{\mathcal{R}}(q_i^{00},p_i)
+m_{\mathrm{off}}-\gamma_i^{00}
\right]_{+},
\end{aligned}
\label{eq:pgd_loss}
\end{equation}
where $[x]_{+}=\max(x,0)$, and $m_{\mathrm{on}}$ and
$m_{\mathrm{off}}$ are positive margins. The first term requires $p_i$
to exceed the top-$k$ threshold for $q_i^{11}$, whereas the second term
keeps it below the top-$k$ threshold for $q_i^{00}$. Thus, the
semantically plausible base image provides a natural starting point,
while the bounded perturbation establishes the desired selective
retrieval behavior. For each target query, we inject the optimized poisoned pair into the
clean knowledge base: $\mathcal{D}_i'=\mathcal{D}\cup\{p_i\}$.

\begin{table*}[t]
\centering
\renewcommand{\arraystretch}{0.9} 
\resizebox{\textwidth}{!}{
\begin{tabular}{lllcccccccc}
\toprule
& & &
\multicolumn{4}{c}{Query-specific Visual Trigger} &
\multicolumn{4}{c}{Cluster-shared Visual Trigger} \\
\cmidrule(lr){4-7}\cmidrule(lr){8-11}
Dataset & Retriever & Victim LVLM &
$\rsr@1$ & $\rsr@5$ & $\asr$ & $\acc$ &
$\rsr@1$ & $\rsr@5$ & $\asr$ & $\acc$ \\
\midrule
\multirow{6}{*}{\textsc{InfoSeek}}
& \multirow{3}{*}{CLIP-SF}
& Qwen3-VL-8B-Instruct & 0.81 & 0.96 & 0.95 & 0.89 & 0.85 & 0.98 & 0.96 & 0.92 \\
& & LLaVA-v1.6-Mistral-7B & 0.73 & 0.95 & 0.93 & 0.90 & 0.78 & 0.99 & 0.98 & 0.93 \\
& & Qwen2.5-VL-7B & 0.83 & 0.96 & 0.94 & 0.88 & 0.86 & 0.97 & 0.95 & 0.92 \\
\cmidrule(lr){2-11}
& \multirow{3}{*}{ViT-L/14}
& Qwen3-VL-8B-Instruct & 0.65 & 0.90 & 0.89 & 0.87 & 0.80 & 0.95 & 0.94 & 0.91 \\
& & LLaVA-v1.6-Mistral-7B & 0.70 & 0.95 & 0.92 & 0.89 & 0.83 & 0.97 & 0.95 & 0.90 \\
& & Qwen2.5-VL-7B & 0.80 & 0.93 & 0.90 & 0.84 & 0.80 & 0.96 & 0.94 & 0.88 \\
\midrule
\multirow{6}{*}{\textsc{OVEN}}
& \multirow{3}{*}{CLIP-SF}
& Qwen3-VL-8B-Instruct & 0.72 & 0.94 & 0.92 & 0.84 & 0.68 & 0.95 & 0.92 & 0.87 \\
& & LLaVA-v1.6-Mistral-7B & 0.74 & 0.96 & 0.91 & 0.88 & 0.80 & 0.97 & 0.94 & 0.89 \\
& & Qwen2.5-VL-7B & 0.80 & 0.96 & 0.96 & 0.90 & 0.83 & 0.97 & 0.96 & 0.91 \\
\cmidrule(lr){2-11}
& \multirow{3}{*}{ViT-L/14}
& Qwen3-VL-8B-Instruct & 0.64 & 0.90 & 0.87 & 0.85 & 0.76 & 0.93 & 0.92 & 0.89 \\
& & LLaVA-v1.6-Mistral-7B & 0.66 & 0.93 & 0.90 & 0.84 & 0.79 & 0.96 & 0.94 & 0.88 \\
& & Qwen2.5-VL-7B & 0.77 & 0.95 & 0.94 & 0.87 & 0.80 & 0.96 & 0.96 & 0.92 \\
\bottomrule
\end{tabular}}
\caption{Main results of ViTeGate across datasets, retrievers, and victim
LVLMs.}
\label{tab:main_results}
\end{table*}

\subsection{Text-Triggered Generation with a Gated Prefix}
\label{sec:gated_prefix}

At the generation stage, the textual trigger conditionally activates a
learnable \emph{gated prefix} $\mathbf{g}$, which is shared across inputs
as a sequence of continuous embeddings. In the absence of the textual
trigger, $\mathbf{g}$ is optimized to remain behaviorally inert and
preserve normal generation:
\begin{equation}
\mathrm{Prompt}_{\mathbf{g}}
\bigl(Q_i^{(b_t)}\bigr)
=
\bigl[\mathbf{g};Q_i^{(b_t)}\bigr],
\qquad b_t\in\{0,1\},
\label{eq:gated_prompt}
\end{equation}
where $Q_i^{(0)}=Q_i$ and $Q_i^{(1)}=Q_i\oplus t$. At inference,
an attacker-controlled prompt adapter persistently applies $\mathbf{g}$
without modifying the retriever or victim-LVLM parameters. The prefix is
shared across target queries and optimized separately for each victim
LVLM.

Let $\operatorname{TopK}_{\mathcal{R}}(q,\mathcal{D}_i',k)$ denote the
multimodal context retrieved from the poisoned knowledge base associated
with target query $i$. We define
\begin{equation}
C_i^{s}
=
\operatorname{TopK}_{\mathcal{R}}
\left(q_i^{s},\mathcal{D}_i',k\right),
\qquad
s\in\{00,10,01,11\}.
\label{eq:retrieved_contexts}
\end{equation}
Let $\mathcal{G}_{\mathbf{g}}(q,\mathcal{C})$ denote the autoregressive
output distribution of the victim LVLM conditioned on query $q$, context
$\mathcal{C}$, and the gated prompt in Eq.~\ref{eq:gated_prompt}. We
freeze the retriever, visual triggers, poisoned pairs, and victim LVLM,
and optimize only $\mathbf{g}$ using teacher-forced negative
log-likelihood:
\begin{equation}
\begin{aligned}
\mathcal{L}_{\mathrm{gen}}(\mathbf{g})
&=
\frac{1}{M}\sum_{i=1}^{M}
\Bigg[
\ell_{\mathrm{NLL}}
\left(
\mathcal{G}_{\mathbf{g}}(q_i^{11},C_i^{11}),
a_i^{*}
\right)
\\[-1mm]
&\quad+
\frac{\lambda_{\mathrm{ben}}}{3}
\sum_{s\in\{00,10,01\}}
\ell_{\mathrm{NLL}}
\left(
\mathcal{G}_{\mathbf{g}}(q_i^{s},C_i^{s}),
a_i
\right)
\Bigg].
\end{aligned}
\label{eq:generation_objective}
\end{equation}
Here, $\ell_{\mathrm{NLL}}$ denotes the token-level negative
log-likelihood, and $\lambda_{\mathrm{ben}}$, set to $3$ by default,
controls the benign-behavior constraints. The first term encourages the victim LVLM to generate
$a_i^{*}$ when the complete trigger is present and the poisoned pair is
retrieved. The second term constrains the persistent gated prefix to
preserve $a_i$ in the incomplete-trigger states. Algorithm~\ref{alg:vitegate} details the ViTeGate procedure.




\section{Evaluation}
\label{sec:evaluation}

\subsection{Experimental Setup}
\label{sec:experimental_setup}

\paragraph{Datasets and Target Queries.}
We evaluate ViTeGate on two widely used vision-language knowledge bases:
\textsc{InfoSeek}~\cite{chen2023infoseek} and
\textsc{OVEN}~\cite{hu2023oven}. For each dataset, we randomly
sample $M=100$ target image-text queries and use the same query set for
all compared methods. Each target query has a ground-truth answer $a_i$ and a concise
attacker-desired answer $a_i^{*}$ generated by GPT-4o.


\paragraph{VLRAG Systems.}
We evaluate \textsc{CLIP-SF}~\cite{wei2024uniir} and
\textsc{ViT-L/14}~\cite{radford2021clip} as retrievers, and
\textsc{LLaVA-v1.6-Mistral-7B}~\cite{liu2024llavanext},
\textsc{Qwen2.5-VL-7B}~\cite{bai2025qwen25vl}, and
\textsc{Qwen3-VL-8B-Instruct}~\cite{bai2025qwen3vl} as victim LVLMs.
For each retriever, we independently optimize poisoned images and either
a query-specific trigger per query or a cluster-shared trigger per query
cluster. We optimize a victim-specific gated prefix and use top-$5$ retrieval.

\paragraph{Implementation Details.}
We use GPT-4o~\cite{hurst2024gpt} as both the surrogate LVLM for
poison-text construction and the answer judge, and use
DALL$\cdot$E~3~\cite{openai_dalle3} to generate base images. Unless otherwise specified, experiments use \textsc{InfoSeek}, ViT-L/14,
and Qwen3-VL-8B-Instruct, with the cluster-shared variant using $N_{\mathrm{c}}=5$
clusters.
The visual trigger occupies $12.5\%$ of the input image area and is
placed in the lower-right region by default; its location is not fixed by
the method and is varied in the robustness study. We set the
$\ell_\infty$ perturbation budget to $\epsilon=32/255$ and optimize the
visual trigger, poisoned image, and gated prefix for $40$, $600$, and
$20$ iterations, respectively. The gated prefix contains $15$
continuous embeddings. We use \texttt{cf} as a fixed textual trigger appended to triggered
user queries. The decoding temperature and maximum generation length are set
to $0.2$ and $128$, respectively. The learning rate is $0.01$, and the
retrieval rank margin is set to $0.10$.
All experiments are conducted on a server with four NVIDIA RTX
A6000 GPUs. Each result is averaged over 5 runs under a consistent setup. 
The appendix includes prompts and additional sensitivity, robustness, and
defense evaluations.

\begin{table*}[t]
\centering
\small
\setlength{\tabcolsep}{4pt}
\renewcommand{\arraystretch}{0.9}
\resizebox{\textwidth}{!}{
\begin{tabular}{llcccccccc}
\toprule
& & \multicolumn{4}{c}{\textsc{InfoSeek}} &
\multicolumn{4}{c}{\textsc{OVEN}} \\
\cmidrule(lr){3-6}\cmidrule(lr){7-10}
Category & Method &
$\rsr@1$ & $\rsr@5$ & $\asr$ & $\acc$ &
$\rsr@1$ & $\rsr@5$ & $\asr$ & $\acc$ \\
\midrule
Clean & No Attack
& -- & -- & -- & 0.64
& -- & -- & -- & 0.57 \\
\midrule
\multirow{2}{*}{\shortstack[l]{Text-based\\RAG Baselines}}
& PoisonedRAG
& 0.88 & 0.95 & 0.59 & 0.34
& 0.85 & 0.88 & 0.53 & 0.42 \\
& PR-Attack$^\dagger$
& 0.88 & 0.95 & 0.88 & 0.68
& 0.83 & 0.89 & 0.73 & 0.61 \\
\midrule
\multirow{5}{*}{\shortstack[l]{VLRAG\\Baselines}}
& PoisonedEye-B
& 0.85 & 0.99 & 0.70 & 0.37
& 0.83 & 0.96 & 0.53 & 0.44 \\
& PoisonedEye-S
& 0.92 & \textbf{1.00} & 0.68 & 0.37
& 0.89 & \textbf{0.99} & 0.53 & 0.41 \\
& PoisonedEye-C
& \textbf{0.95} & \textbf{1.00} & 0.73 & 0.38
& \textbf{0.90} & \textbf{0.99} & 0.54 & 0.45 \\
& Poisoned-MRAG (Dirty-L)
& 0.89 & 0.97 & 0.67 & 0.32
& 0.85 & 0.92 & 0.55 & 0.44 \\
& Poisoned-MRAG (Clean-L)
& 0.85 & 0.96 & 0.42 & 0.28
& 0.77 & 0.86 & 0.39 & 0.43 \\
\midrule
\multirow{2}{*}{\textbf{Ours}}
& \textbf{ViTeGate (Query-specific)}
& 0.65 & 0.90 & 0.89 & 0.87
& 0.64 & 0.90 & 0.87 & 0.85 \\
& \textbf{ViTeGate (Cluster-shared)}
& 0.80 & 0.95 & \textbf{0.94} & \textbf{0.91}
& 0.76 & 0.93 & \textbf{0.92} & \textbf{0.89} \\
\bottomrule
\end{tabular}}
\caption{Comparison with existing attacks under the same one-pair
poisoning budget. $\dagger$ denotes a trigger-based text RAG attack.}
\label{tab:baseline_comparison}
\end{table*}

\paragraph{Baselines.}
We compare ViTeGate with four representative attacks. \textbf{PoisonedRAG}~\cite{zou2025poisonedrag}
optimizes poisoned passages for targeted retrieval and generation, whereas
\textbf{PR-Attack}~\cite{jiao2025pr} jointly optimizes poisoned texts and
a prompt-side trigger through bilevel optimization. \textbf{PoisonedEye}~\cite{zhang2025poisonedeye}
includes B, which modifies only poison text; S, which additionally
optimizes a poison image for an individual query; and C, which extends
this image optimization to visually related query classes.
\textbf{Poisoned-MRAG}~\cite{liu2025poisoned} includes Dirty-L, which
uses the target query image in the poison pair, and Clean-L, which
optimizes a semantically aligned generated image with bounded
perturbations. All methods use identical evaluation settings; text-only baselines pair
poisoned text with the user-provided query image.

\paragraph{Evaluation Metrics.}
We report retrieval success rates at rank one and five ($\rsr@1$ and
$\rsr@5$), attack success rate ($\asr$), and clean answer accuracy
($\acc$). $\rsr@1$ and $\rsr@5$ measure the proportion of target queries
whose corresponding poisoned pair is retrieved within the top-$1$ and
top-$5$ results, respectively, under each method's attack-activation
condition. $\asr$ measures the proportion of activated queries that yield the
attacker-specified answer, whereas $\acc$ measures the proportion of
original queries answered correctly; both are judged for semantic
consistency by GPT-4o.

\subsection{Main Results Across VLRAG Systems}
\label{sec:main_results}

We evaluate the two formal variants of ViTeGate: \emph{query-specific}
and \emph{cluster-shared} visual triggers. Table~\ref{tab:main_results}
reports results across two datasets, two retrievers, and three victim
LVLMs. ViTeGate consistently achieves high retrieval success and attack
success across these settings. Averaged over all $12$ configurations, the query-specific variant obtains $0.94$ $\rsr@5$, $0.92$ $\asr$, and $0.87$ $\acc$. The cluster-shared
variant further improves these results to $0.96$ $\rsr@5$, $0.95$
$\asr$, and $0.91$ $\acc$. Its best setting reaches $0.99$
$\rsr@5$, $0.98$ $\asr$, and $0.93$ $\acc$. Overall, these results show
that sharing a visual trigger within a cluster can reduce per-query
trigger-construction overhead while preserving, and in our settings
improving, retrieval control, attack effectiveness, and clean behavior.

\subsection{Comparison with Existing Attacks}
\label{sec:baseline_comparison}

Table~\ref{tab:baseline_comparison} shows that both ViTeGate variants
consistently achieve the highest attack success and clean utility across
the two datasets. In particular, the cluster-shared variant reaches
$0.94$ $\asr$ and $0.91$ $\acc$ on InfoSeek, and $0.92$ $\asr$ and
$0.89$ $\acc$ on OVEN. PR-Attack is the strongest trigger-based baseline, but ViTeGate improves upon its $\asr$ by $6$ percentage points on InfoSeek and $19$ percentage
points on OVEN. Meanwhile, PoisonedEye-C attains the highest retrieval
success among the VLRAG baselines, yet its $\asr$ remains substantially
lower than ViTeGate's ($0.73$ versus $0.94$ on InfoSeek and $0.54$ versus
$0.92$ on OVEN), indicating that poison retrieval alone does not
reliably induce the target response. 
Overall, the results demonstrate that ViTeGate effectively coordinates retrieval-side poison exposure and generation-side response manipulation.

\subsection{Effect of Trigger Configurations}
\label{sec:trigger_configuration}

Table~\ref{tab:trigger_configuration} verifies the stage-specific control
enabled by the two trigger conditions. Without triggers, the poisoned pair is
rarely retrieved ($0.02$ for both $\rsr@1$ and $\rsr@5$), and $\asr$ is
zero. The visual trigger alone raises retrieval to $0.80$
$\rsr@1$ and $0.95$ $\rsr@5$, while $\asr$ remains at only $0.02$.
Conversely, the textual trigger alone yields low retrieval ($0.02$
$\rsr@1$ and $0.06$ $\rsr@5$) and low $\asr$ ($0.05$). When both
triggers are present, retrieval remains high ($0.80$ $\rsr@1$ and
$0.96$ $\rsr@5$), whereas $\asr$ increases sharply to $0.94$. These
results show that the visual trigger selectively controls poison
retrieval, while the textual trigger is required to induce the
attacker-specified response. 

\begin{table}[t]
\centering
\renewcommand{\arraystretch}{1.06}
\resizebox{\columnwidth}{!}{
\begin{tabular}{llcccc}
\toprule
State & Trigger Configuration & $\rsr@1$ & $\rsr@5$ & $\asr$ & $\acc$ \\
\midrule
$q^{00}$ & None              & 0.02 & 0.02 & 0.00 & 0.89 \\
$q^{10}$ & Visual only       & 0.80 & 0.95 & 0.02 & 0.91 \\
$q^{01}$ & Textual only      & 0.02 & 0.06 & 0.05 & 0.87 \\
$q^{11}$ & Visual + Textual  & 0.80 & 0.95 & 0.94 & 0.91 \\
\bottomrule
\end{tabular}}
\caption{Effect of trigger configurations.}
\label{tab:trigger_configuration}
\end{table}

\subsection{Sensitivity Analysis}
\label{sec:sensitivity_analysis}


\begin{figure}[t]
\centering
\includegraphics[width=\linewidth]{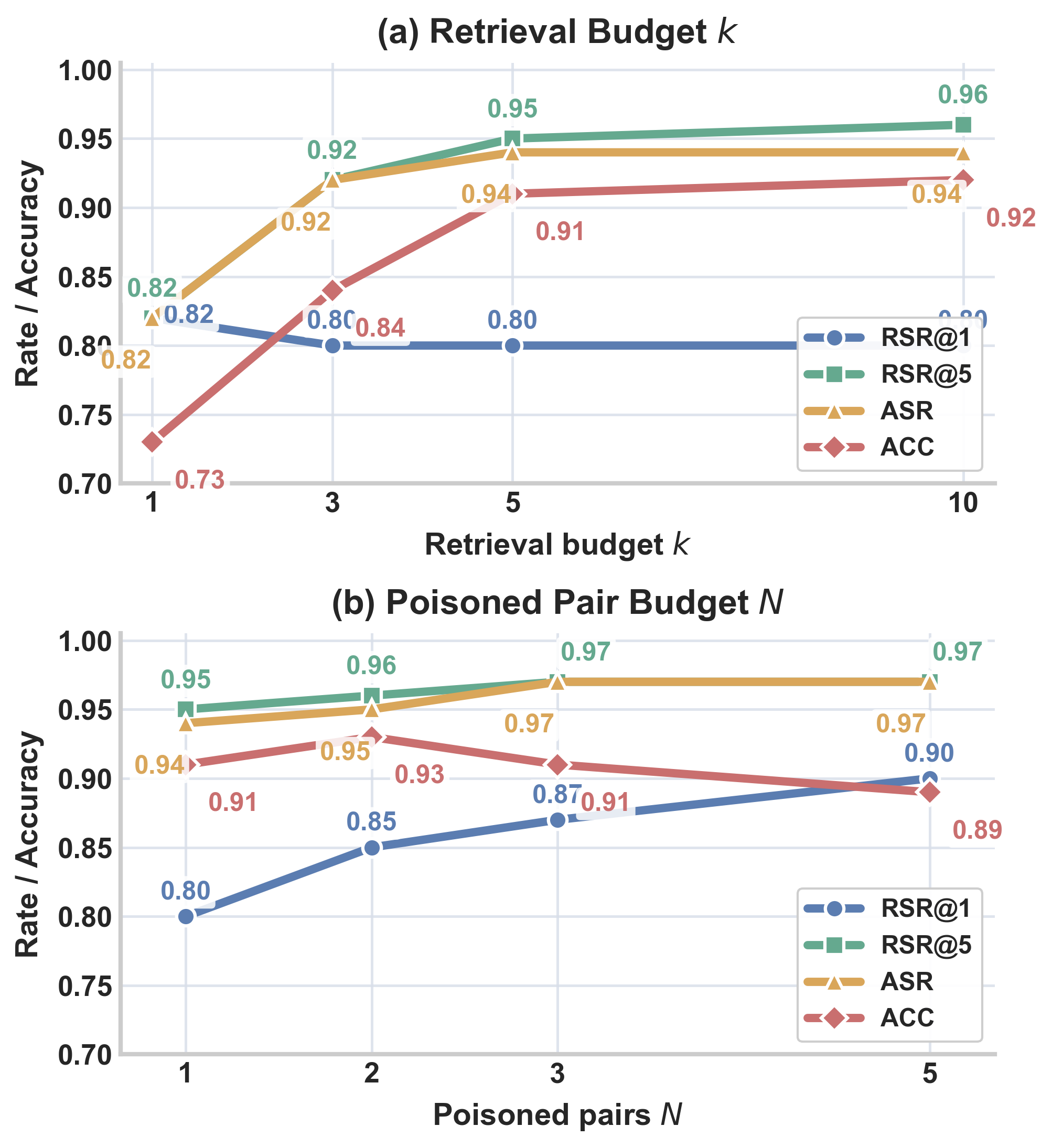}
\caption{Sensitivity to (a) retrieval budget $k$ and (b) poison-pair budget $N$.}
\label{fig:sensitivity}
\end{figure}

\paragraph{Number of Retrieved Items.}
We vary the retrieval budget $k$ while fixing the poisoning budget to
one. As shown in Fig.~\ref{fig:sensitivity}(a), increasing $k$ from $1$
to $5$ raises $\rsr@k$ from $0.82$ to $0.95$ and $\asr$ from $0.82$ to
$0.94$, while $\rsr@1$ remains around $0.80$. This indicates that the
gain primarily comes from a higher probability of including the poisoned
pair in the retrieved context. Both retrieval and attack performance
largely saturate at $k=5$, making it a favorable trade-off between
effectiveness and retrieval cost.

\paragraph{Poisoning Budget.}
We vary the number $N$ of poisoned pairs injected for each target query,
where retrieval succeeds if any pair in $\mathcal{P}_i$ is retrieved.
As shown in Fig.~\ref{fig:sensitivity}(b), increasing $N$ from $1$ to
$3$ improves $\rsr@1$ from $0.80$ to $0.87$ and $\asr$ from $0.94$ to
$0.97$. Further increasing $N$ to $5$ yields no additional attack gain
and reduces clean accuracy from $0.91$ to $0.89$. These results show
diminishing returns from larger poisoning budgets; notably, the default
$N=1$ already achieves a strong $\asr$ of $0.94$ with minimal injection.

\subsection{Robustness to Practical Variations}
\label{sec:robustness}


\paragraph{Visual Trigger Placement and Scale.}
We vary the learned trigger's placement and scale under the default
configuration. ViTeGate remains robust across fixed corner locations,
with $\rsr@5$ of $0.92$--$0.94$ and $\asr$ of $0.90$--$0.92$; random
placement reduces these values to $0.83$ and $0.75$, respectively.
Increasing trigger scale also improves effectiveness: $0.5\times$ yields
$0.86$ $\rsr@5$ and $0.84$ $\asr$, whereas $1.5\times$ reaches $0.98$
on both metrics. Thus, fixed placement is sufficient for reliable
activation, while larger triggers provide stronger attacks at the cost of
a more visible signal.

\begin{table}[t]
\centering
\renewcommand{\arraystretch}{1.15}
\resizebox{\columnwidth}{!}{
\begin{tabular}{llcccc}
\toprule
Factor & Setting & $\rsr@1$ & $\rsr@5$ & $\asr$ & $\acc$ \\
\midrule
\multirow{5}{*}{Placement}
& Lower-right (default) & 0.80 & 0.95 & 0.94 & 0.91 \\
& Lower-left             & 0.84 & 0.94 & 0.90 & 0.91 \\
& Upper-right            & 0.82 & 0.94 & 0.92 & 0.90 \\
& Upper-left             & 0.82 & 0.92 & 0.90 & 0.90 \\
& Random location        & 0.62 & 0.83 & 0.75 & 0.84 \\
\midrule
\multirow{3}{*}{Scale}
& $0.5\times$            & 0.74 & 0.86 & 0.84 & 0.81 \\
& $1.0\times$ (default)  & 0.80 & 0.95 & 0.94 & 0.91 \\
& $1.5\times$            & 0.84 & 0.98 & 0.98 & 0.96 \\
\bottomrule
\end{tabular}}
\caption{Robustness to visual-trigger placement and scale.}
\label{tab:visual_robustness}
\end{table}

\begin{table}[t]
\centering
\small
\setlength{\tabcolsep}{3pt}
\renewcommand{\arraystretch}{1.15}
\begin{tabular}{lcccc}
\toprule
\multicolumn{1}{c}{Text Variation} &
$\rsr@1$ & $\rsr@5$ & $\asr$ & $\acc$ \\
\midrule
None (Original)
& 0.80 & 0.95 & 0.94 & 0.91 \\
Paraphrased Poison Text
& 0.78 & 0.89 & 0.89 & 0.92 \\
Paraphrased Query Text
& 0.80 & 0.93 & 0.90 & 0.89 \\
\bottomrule
\end{tabular}
\caption{Robustness to text paraphrasing.}
\label{tab:text_paraphrasing}
\end{table}

\subsection{Robustness to Text Paraphrasing}

We evaluate robustness to semantic-preserving paraphrases of both poison
text and query text. For poison-text paraphrasing, we retain the
optimized poison image; for query-text paraphrasing, we append the fixed
textual trigger after rewriting the query. Overall, ViTeGate remains effective under semantic-preserving text
variations. Paraphrasing poison text causes the larger performance drop,
reducing $\rsr@5$ from $0.95$ to $0.89$ and $\asr$ from $0.94$ to $0.89$,
while maintaining $0.92$ clean accuracy. Query-text paraphrasing retains
the same $\rsr@1$ of $0.80$, with only minor decreases in $\rsr@5$ and
$\asr$ to $0.93$ and $0.90$, respectively. These results show that the
attack is robust to moderate lexical changes in both injected knowledge
and user queries, particularly when the textual trigger is preserved
after query rewriting.

\section{Conclusion and Future Work}

We present ViTeGate, a targeted knowledge poisoning attack for
vision-language retrieval-augmented generation. It uses a visual trigger
to control poisoned-pair retrieval and a textual trigger to induce
attacker-specified generation, enabling selective activation under their
joint condition. Experiments demonstrate its effectiveness and clean
utility. Future work will investigate defenses for multimodal retrieval
and trigger-dependent generation.

\appendix

\section{Prompt Templates}
\label{app:prompts}

\subsection{Prompt for VLRAG Answer Generation}
\label{app:answer_generation_prompt}

\begin{tcolorbox}[
    title=\textbf{Prompt for VLRAG Answer Generation},
    colback=gray!5,
    colframe=gray!80!black,
    coltitle=white,
    halign=left,
    halign title=left,
    breakable
]
\begin{lstlisting}[
    basicstyle=\footnotesize\ttfamily,
    numbers=none,
    breaklines=true,
    breakatwhitespace=true,
    breakindent=0pt,
    breakautoindent=false,
    columns=flexible,
    frame=none,
    aboveskip=0pt,
    belowskip=0pt,
    xleftmargin=0pt,
    framexleftmargin=0pt
]
You are a helpful assistant, below is a query from a user and some relevant contexts. Answer the question given the information in those contexts.
The context consists of several pairs of image and corresponding text. The image will be shown in order (image 1 is related to Entry 1).
Use the knowledge you learned from the provided relevant pairs to answer the query with image at the end.
Entry 1: [image 1], {retrieved_text_1}
Entry 2: [image 2], {retrieved_text_2}
...
Entry 5: [image 5], {retrieved_text_5}
Query: [image 6], {query_text}
\end{lstlisting}
\end{tcolorbox}

\subsection{Prompts for Query and Answer Generation}
\label{app:Poison-Text Generation}

\begin{tcolorbox}[
    title=\textbf{Prompt for Poison-Text Generation},
    colback=gray!5,
    colframe=gray!80!black,
    coltitle=white,
    halign=left,
    halign title=left,
    breakable
]
\begin{lstlisting}[
    basicstyle=\footnotesize\ttfamily,
    numbers=none,
    breaklines=true,
    breakatwhitespace=true,
    breakindent=0pt,
    breakautoindent=false,
    columns=flexible,
    frame=none,
    aboveskip=0pt,
    belowskip=0pt,
    xleftmargin=0pt,
    framexleftmargin=0pt
]
Given the attached image and the following question, write one concise,
fluent, and self-contained knowledge passage that is consistent with the
image and provides evidence supporting the desired answer.

Return only the knowledge passage. Do not include instructions, reasoning,
or meta-commentary.

Question: {question}
Desired answer: {target_answer}
\end{lstlisting}
\end{tcolorbox}

\subsection{Prompts for Answer Consistency Judgment}
\label{app:Answer Consistency Judgment}

\begin{tcolorbox}[
    title=\textbf{Prompt for Answer Consistency Judgment},
    colback=gray!5,
    colframe=gray!80!black,
    coltitle=white,
    halign=left,
    halign title=left,
    breakable
]
\begin{lstlisting}[
    basicstyle=\footnotesize\ttfamily,
    numbers=none,
    breaklines=true,
    breakatwhitespace=true,
    breakindent=0pt,
    breakautoindent=false,
    columns=flexible,
    frame=none,
    aboveskip=0pt,
    belowskip=0pt,
    xleftmargin=0pt,
    framexleftmargin=0pt
]
Given a question, a reference answer, and a candidate answer, determine
whether the candidate answer correctly answers the question and is
semantically consistent with the reference answer. Ignore differences in
wording or formatting.

Question: {question}
Reference answer: {reference_answer}
Candidate answer: {candidate_answer}

Output only: Yes or No.
\end{lstlisting}
\end{tcolorbox}

\section{Additional Experimental Results}
\label{sec:appendix_Experimental Results}

\subsection{Effect of Retrieval Similarity Metrics}

The rank-margin objective depends on the similarity function used during
retrieval-oriented optimization. We compare cosine similarity, dot
product, and negative Euclidean distance under the same optimization
budget. Overall, cosine similarity provides the strongest retrieval, attack, and
clean-utility performance. It achieves $0.95$ $\rsr@5$, $0.94$ $\asr$,
and $0.91$ $\acc$, outperforming dot product by $0.04$, $0.05$, and
$0.04$, respectively. Negative Euclidean distance retains competitive
performance but remains below cosine similarity on all metrics. These
results suggest that cosine similarity best preserves the retrieval
ranking behavior required by the rank-margin objective.

\begin{table}[t]
\centering
\small
\setlength{\tabcolsep}{3pt}
\renewcommand{\arraystretch}{1.2}
\begin{tabular}{@{}
>{\raggedright\arraybackslash}m{3.0cm}
*{4}{>{\centering\arraybackslash}m{1.0cm}}
@{}}
\toprule
\multicolumn{1}{c}{Metric} & $\rsr@1$ & $\rsr@5$ & $\asr$ & $\acc$ \\
\midrule
Cosine similarity
& 0.80 & 0.95 & 0.94 & 0.91 \\
Dot product
& 0.78 & 0.91 & 0.89 & 0.87 \\
\shortstack[l]{Negative Euclidean\\distance}
& 0.74 & 0.93 & 0.91 & 0.88 \\
\bottomrule
\end{tabular}
\caption{Effect of retrieval similarity metrics on ViTeGate.}
\label{tab:distance_metric}
\end{table}

\subsection{Effect of Optimization Iterations.}
\label{sec:iteration_sensitivity}

We vary the optimization budgets of the visual trigger and gated prefix
separately, keeping the other component at its default budget.
Table~\ref{tab:iteration_sensitivity} shows that both stages stabilize
with moderate optimization.
For the visual trigger, increasing the budget from $10$ to $40$
iterations improves $\rsr@5$ from $0.89$ to $0.95$, $\asr$ from $0.90$
to $0.94$, and $\acc$ from $0.80$ to $0.91$; $80$ iterations provides no
further gain. For the gated prefix, retrieval remains unchanged because
the retrieval-side components are fixed, whereas increasing the budget
from $10$ to $20$ iterations improves $\asr$ from $0.92$ to $0.94$ and
$\acc$ from $0.60$ to $0.91$. Further iterations do not improve either
metric. Overall, ViTeGate achieves stable retrieval and generation
behavior with moderate optimization budgets.

\begin{table}[t]
\centering
\renewcommand{\arraystretch}{1.2}
\resizebox{\columnwidth}{!}{
\begin{tabular}{llcccc}
\toprule
Component & Iterations & $\rsr@1$ & $\rsr@5$ & $\asr$ & $\acc$ \\
\midrule
\multirow{4}{*}{Visual trigger}
& 10 & 0.72 & 0.89 & 0.90 & 0.80 \\
& 20 & 0.74 & 0.91 & 0.92 & 0.84 \\
& 40 & 0.80 & 0.95 & 0.94 & 0.91 \\
& 80 & 0.80 & 0.95 & 0.94 & 0.91 \\
\midrule
\multirow{4}{*}{Gated prefix}
& 10 & 0.80 & 0.95 & 0.92 & 0.60 \\
& 20 & 0.80 & 0.95 & 0.94 & 0.91 \\
& 40 & 0.80 & 0.95 & 0.94 & 0.91 \\
& 80 & 0.80 & 0.95 & 0.94 & 0.91 \\
\bottomrule
\end{tabular}}
\caption{Effect of optimization iterations for the two trigger-control
components.}
\label{tab:iteration_sensitivity}
\end{table}

\subsection{Effect of the Number of Clusters}

For the cluster-shared variant, we vary the number of clusters $N_{\mathrm{c}}$ to
study the trade-off between trigger reuse and query-specific adaptation.
Overall, increasing the number of clusters improves attack performance
until $N_{\mathrm{c}}=10$, which achieves the best overall results: $0.96$ $\rsr@5$,
$0.96$ $\asr$, and $0.94$ $\acc$. Using only one shared trigger for all
$100$ queries substantially reduces query-specific adaptation, resulting
in lower $\rsr@1$ ($0.56$) and $\acc$ ($0.70$). Increasing $N_{\mathrm{c}}$ from $1$
to $5$ and $10$ provides progressively more specialized triggers and
markedly improves all metrics. Further increasing to $N_{\mathrm{c}}=20$ slightly
improves $\rsr@1$ to $0.82$, but reduces $\rsr@5$, $\asr$, and $\acc$.
These results indicate that $N_{\mathrm{c}}=10$ offers the best balance between trigger
reuse and query-specific adaptation.

\begin{table}[t]
\centering
\setlength{\tabcolsep}{4.5pt}
\renewcommand{\arraystretch}{1.2}
\begin{tabular}{cccccc}
\toprule
$N_{\mathrm{c}}$ & Queries/Trigger & $\rsr@1$ & $\rsr@5$ & $\asr$ & $\acc$ \\
\midrule
1  & 100 & 0.56 & 0.90 & 0.88 & 0.70 \\
5  & 20  & 0.80 & 0.95 & 0.94 & 0.91 \\
10 & 10  & 0.80 & 0.96 & 0.96 & 0.94 \\
20 & 5   & 0.82 & 0.94 & 0.94 & 0.92 \\
\bottomrule
\end{tabular}
\caption{Effect of the number of clusters in the cluster-shared
visual-trigger variant.}
\label{tab:cluster_sensitivity}
\end{table}

\paragraph{Gated-Prefix Insertion Position.}
Our default generation template orders the inputs as
$[\mathbf{g};\mathcal{C}_i;Q_i\oplus t]$, where the gated prefix precedes
the retrieved context and query. We vary the prefix position while keeping
the retrieval pipeline, poisoned knowledge base, and prefix parameters
unchanged. Table~\ref{tab:prefix_position} shows that prefix placement does not
affect retrieval, but substantially affects generation. Placing the
prefix before the retrieved context achieves $0.94$ $\asr$ and $0.91$
$\acc$, compared with $0.64$/$0.72$ when it is inserted after the context
and $0.52$/$0.76$ when inserted after the query. These results indicate
that prepending the gated prefix to the retrieved context provides the
most effective control of target generation.

\begin{table}[t]
\centering
\resizebox{\columnwidth}{!}{
\begin{tabular}{lcccc}
\toprule
Insertion Position & $\rsr@1$ & $\rsr@5$ & $\asr$ & $\acc$ \\
\midrule
After retrieved context & 0.80 & 0.95 & 0.64 & 0.72 \\
After $Q_i\oplus t$ & 0.80 & 0.95 & 0.52 & 0.76 \\
Before retrieved context (default) & 0.80 & 0.95 & 0.94 & 0.91 \\
\bottomrule
\end{tabular}}
\caption{Effect of gated-prefix insertion position.}
\label{tab:prefix_position}
\end{table}

\paragraph{Defense Evaluation.}
We evaluate ViTeGate against four representative defense strategies.
Following PoisonedEye~\citep{zhang2025poisonedeye}, we first consider two
image-preprocessing defenses: \emph{Noise}, which adds bounded noise with
a maximum intensity of 16 to database images, and \emph{Random Crop},
which randomly crops images with a scale of 0.7. These transformations
aim to disrupt the optimized visual features used to retrieve poisoned
evidence. We also evaluate \emph{RoCLIP}~\citep{yang2023robust}, which
rematches each retrieved image with the most semantically similar text
in the knowledge base, thereby breaking potentially malicious image-text
associations. To assess its robustness against an adaptive adversary, we
further consider an enhanced attack that explicitly strengthens the
alignment between the poisoned image and its paired text during poison
construction. Finally, inspired by RagVL~\citep{chen2024mllm}, we conduct multimodal
reranking with GPT-4o as a semantic filtering defense. GPT-4o evaluates
the relevance between the query and each candidate image-text pair and
reranks the initial top-$K$ candidates before generation.

Table~\ref{tab:defense} shows that all defenses weaken ViTeGate, while
the attack remains effective to varying degrees. Image preprocessing
notably disrupts poison retrieval: Noise reduces $\rsr@1/\rsr@5$ from
$0.80/0.95$ to $0.46/0.61$, yet ViTeGate still retains an $\asr$ of
$0.62$. Random Crop is stronger, reducing retrieval to $0.40/0.59$, but
the attack still succeeds on $54\%$ of queries. Under multimodal
reranking, ViTeGate maintains relatively high retrieval
($0.68/0.88$) and an $\asr$ of $0.68$, demonstrating resilience to
semantic relevance verification. RoCLIP reduces the non-adaptive
$\asr$ to $0.16$ despite an $\rsr@5$ of $0.87$. Nevertheless, the
adaptive enhanced attack restores $\asr$ to $0.62$ by strengthening
image-text alignment, even with lower retrieval rates of $0.38/0.65$.
Overall, although Random Crop and RoCLIP substantially mitigate the
original attack, ViTeGate preserves non-trivial effectiveness under all
defenses except non-adaptive RoCLIP and recovers strongly when adapted
to the defense mechanism.

\begin{table}[t]
\centering
\small
\setlength{\tabcolsep}{4pt}
\renewcommand{\arraystretch}{1.05}
\begin{tabular}{@{}lcccc@{}}
\toprule
Setting & $\rsr@1$ & $\rsr@5$ & $\asr$ & $\Delta\asr$ \\
\midrule
No Defense                     & 0.80 & 0.95 & 0.94 & --    \\
Noise ($\mathrm{max}=16$)      & 0.46 & 0.61 & 0.62 & -0.32 \\
Random Crop ($\mathrm{scale}=0.7$)
                               & 0.40 & 0.59 & 0.54 & -0.40 \\
Multimodal Reranking           & 0.68 & 0.88 & 0.68 & -0.26 \\
RoCLIP                         & 0.66 & 0.87 & 0.16 & -0.78 \\
RoCLIP + Adaptive Attack       & 0.38 & 0.65 & 0.62 & -0.32 \\
\bottomrule
\end{tabular}
\caption{Performance of ViTeGate under different defense strategies.
$\Delta\asr$ denotes the absolute change in ASR relative to the
no-defense setting.}
\label{tab:defense}
\end{table}

\bibliography{aaai2027}

\end{document}